\documentclass[showpacs, prb,twocolumn,preprintnumbers , superscriptaddress, aps]{revtex4-2}

\usepackage{color}
\usepackage{amsmath,amssymb}
\usepackage{pifont}
\usepackage{amssymb}  
\usepackage{bbold}
\usepackage{float}
\usepackage{stfloats}
\usepackage[normalem]{ulem}

\usepackage[labelfont=bf]{caption} 
\usepackage{tikz}
\usepackage{bbold}
\usepackage{makecell}
\usepackage{pifont}   
\usepackage{graphicx} 
\usepackage{dcolumn}  
\usepackage{bm}       
\usepackage{multirow} 
\usepackage{placeins}
\usepackage[colorlinks]{hyperref}
\usepackage{mathtools}
\usepackage{subfigure}
\usepackage{amsmath}  
\newcommand\scalemath[2]{\scalebox{#1}{\mbox{\ensuremath{\displaystyle #2}}}}

\usepackage[normalem]{ulem}

\newcommand{\vect}[1]{\mathbf{#1}}

\usepackage{breqn}
\usepackage{mathrsfs}

\usepackage{multirow}

\begin{document}

\title{Variational principle for the time evolution operator, its usefulness in effective theories of condensed matter systems and a glimpse into the role played by the quantum geometry of unitary transformations}

\author{Michael Vogl}
		\affiliation{Physics Department$,$
		King Fahd University
		of Petroleum $\&$ Minerals$,$
		Dhahran 31261$,$ Saudi Arabia}
    \affiliation{Interdisciplinary Research Center (IRC) for Intelligent Secure Systems$,$ KFUPM$,$ Dhahran$,$ Saudi Arabia}
\date{\today}
	\begin{abstract}
This work discusses a variational approach to determining the time evolution operator. We directly see a glimpse of how a generalization of the quantum geometric tensor for unitary operators plays a central role in parameter evolution. We try the method with the simplest ansatz (a power series in a time-independent Hamiltonian), which yields considerable improvements over a Taylor series. These improvements are because, unlike for a Taylor series of $\exp(-iHt)$, time $t$ is not forced to appear in the same order as $H$, giving more flexibility for the description. We demonstrate that our results can also be employed to improve degenerate perturbation theory in a non-perturbative fashion. We concede that our approach described here is most useful for finite-dimensional Hamiltonians. As a first example of applications to perturbation theory, we present AB bilayer graphene, which we downfolded to a 2x2 model; our energy results considerably improve typical second-order degenerate perturbation theory. We then demonstrate that the approach can also be used to derive a non-perturbatively valid Heisenberg Hamiltonian. Here, the approach for a finite-size lattice yields excellent results. However, the corrections are not ideal for the thermodynamic limit (they depend on the number of sites $N$). Nevertheless, the approach adds almost no additional technical complications over typical perturbative expansions of unitary operators, making it ready for deployment in physics questions. One should expect considerably improved couplings for the degenerate perturbation theory of finite-size systems. More work is needed in the many-body case, and we suggest a possible remedy to issues with the thermodynamic limit. Our work hints at how the appearance of mathematically beautiful concepts like quantum geometry can indicate an opportunity to dig for improved approximations beyond typical perturbation theory.
 \end{abstract}

\maketitle

\section{Introduction}
Much of modern theoretical physics is built on only a few key concepts. Most predictions are based on what is, in essence, solutions of harmonic oscillators \cite{polyakov1987gauge}, effective (low dimensional) theories (in the words of Xiao-Gang Wen if [one] wants to remember only one thing from field theory, then remember effective theory \cite{10.1093/acprof:oso/9780199227259.001.0001}), perturbative approaches \cite{simon1991fifty,suzuki1983degenerate,RevModPhys.44.602}, a zoo of mean-field theories \cite{PhysRevB.108.155109,PhysRevLett.133.150401} and variational principles \cite{TILLY20221,Yuan_2019,Cerezo_2021}. In the past decades, exponentially growing computational resources have permitted us to push these approaches to ever higher levels of sophistication that permit the treatment of systems with growing levels of complexity. For instance, one of the most sophisticated modern variants of mean-field theory is dynamical mean field theory (DMFT) \cite{RevModPhys.68.13,RevModPhys.78.865,RevModPhys.86.779}, which allows a convenient treatment of the Hubbard model. Another development that was heavily inspired by mean-field theory is density functional theory (DFT) \cite{RevModPhys.87.897,PhysRev.140.A1133,orio2009density,Mardirossian02102017}, which is the workhorse of condensed matter theory when trying to obtain experimentally relevant parameters for simplified theories and much more. Modern developments that spawned from variational principles are various tensor network methods \cite{RevModPhys.93.045003,annurevtensnetwork,ORUS2014117} and newly developed variational neural network methods \cite{PhysRevLett.133.106701,PhysRevB.110.195131,PhysRevLett.122.250503,PhysRevB.106.165111,PhysRevLett.131.081601}, which are starting to leave a lasting imprint on the landscape of numerical methods for the solution of quantum systems. 

In addition to methods for solving physical problems, the principal driving force for many deep analytical insights in physics has always been observations about mathematical beauty. For instance, Noether's observations \cite{noether1983invariante} about the connection between symmetries and conservation laws are, in essence, what culminated in the development of the standard model of particle physics \cite{RevModPhys.71.S96,weinberg2004making}. For a long time, the Landau paradigm of symmetry breaking \cite{PhysRevResearch.2.023031,PhysRevB.70.144407} gave us a relatively coherent picture of phase transitions. This idea was expanded upon later when physicists took note of the beauty of topologically protected phases of matter such as topological insulators \cite{RevModPhys.82.3045,RevModPhys.83.1057,moore2010birth} or even more recently topologically protected quantum bits \cite{KITAEV20062,gladchenko2009superconducting,ioffe2002topologically}. Similarly, geometry has always played an important role in physics from the earliest analysis of shapes of planetary orbits \cite{boccaletti2001epicyclesgreekskeplersellipse} to the geometrical interpretation of general relativity \cite{carroll2019spacetime}. More recently, quantum geometry of wavefunctions has shown up as an important contribution in response functions for experimental quantities \cite{PhysRevB.110.245122,PhysRevLett.131.240001} and here the trend continues.

We are inspired by the previous progress mentioned above, and the paper aims to make a small contribution to the growing literature on approximate solution methods for physical problems. To varying extents, the work builds on all of the above-mentioned ideas. In particular, we will use a variational principle for the time evolution operator and give a glimpse into how it links to ideas from quantum geometry. We then apply it directly to understand better how to approximate time evolution operators. Motivated here by encouraging results for the accuracy of the approximate time evolution, we apply the method in deriving effective low-energy Hamiltonians - beyond a perturbative limit. Our work shows how quantum geometry could leave its imprints on non-perturbative approximations of low-energy effective Hamiltonians. Our work underlines the importance of the idea that mathematical beauty, even in the context of approximations, can inform progress. 

\section{Variational approach to the time evolution operator}
We start our discussion by noting that the Schr\"odinger equation for the time evolution operator
\begin{equation}
    i\partial_t U=HU
\end{equation}
can be derived by the following action
\begin{equation}
    S=\int dt \mathrm{tr}[U^\dagger (i\partial_t U-HU)]
    \label{eq:action}
\end{equation}
if we vary with respect to $u^*_{ij}$ (the matrix elements of $U^\dagger$).
Like typical in a variational principle, we may want to parametrize an approximate solution $U$ with parameters $c_i(t)$ and then vary with respect to $c_i^*$ such that we obtain a set of equations for the variational parameters, which is given below
\begin{equation}
i\sum_k\mathrm{tr}\left[\left(\frac{\partial U}{\partial c_j}\right)^\dagger\frac{\partial U}{\partial c_k}\right]\partial_t c_k-\mathrm{tr}\left[\left(\frac{\partial U}{\partial c_j}\right)^\dagger H U\right]=0.
\end{equation}
We note in passing that the first term $\mathrm{tr}\left[\left(\partial_{c_j} U\right)^\dagger\partial_{c_k} U\right]$ is a generalization of the naive quantum geometric tensor - known from wavefunctions - to the case of unitary operators. That is, if similar to the discussion in \cite{cheng2013quantumgeometrictensorfubinistudy}, we compute the quantum distance $\| U(c+dc)-U(c)\|^2$ to lowest non-vanishing order in $dc$, we obtain the interpretation as a metric tensor (the norm chosen here is the Frobenius norm). Of course, it is important to note that a change by a $U(1)$ transformation in physical cases should have no impact. To obtain an actual physical tensor, one has to ensure gauge invariance - in the case of a system with gauge or other symmetries, one should ensure something similar. This observation gives us a glimpse into the role of the quantum geometric tensor in all that follows. However, we will not explicitly follow down this line of reasoning; instead, we will seek the simplest possible solutions to the equation. We choose this approach to ensure that the treatments of the main goals of the paper stay as lucid as possible.

In what follows, we consider only the simplest case of a Hamiltonian $H^\dag=H$ that is hermitian and time-independent. Moreover, we will make the simplest possible ansatz 
\begin{equation}
    U=\sum_{j}^{n^*}c_j(t) H^j,
\end{equation}
where we introduced a cut-off order $n^*$. It is important to note that such an ansatz should often be expected to yield better results than a simple Taylor expansion of a matrix exponential. For instance, it will yield exact results whenever $H^{n^*+1}=\sum_i^{n^*} c_i H^i$. This relation holds, for instance, whenever the dimension of the Hilbert space $d=n^*$ or the Hamiltonian is an element of the $Z^{n^*}$ group, i.e. $H^{n^*}=\mathbb{1}$.
After a brief computation, we find that we have $n^*$ equations
\begin{equation}
\sum_k \left[i\partial_t c_k\mathrm{tr}(H^{k+l})  -c_k\mathrm{tr}(H^{k+l+1}) \right]=0
\end{equation}
for our coefficients $c_k(t)$.

\section{Properties of the expansion that is polynomial in $H$}
 Because it is often not practical to work with high powers of a Hamiltonian, we consider the simple case of $n^*=2$ first. It serves as a first step to understanding our polynomial expansion's properties. Besides being the most practical case, it is the most relevant to degenerate second-order perturbation theory - our later application. To get useful insights into how the results of our approach exactly improve on a typical Taylor series, we make a comparison. That is, below, we compare the so-called $l_2$ distance 
\begin{equation}
\begin{aligned}
    l_2(t)&=\frac{1}{2 \sqrt{D_{\mathrm{dim}}}}\left\|U_{\mathrm{ex}}(t)-U_{\mathrm{a}}(t)\right\|_{\mathrm{Frob}} \\ \|A\|_{\mathrm{Frob}}&=\sqrt{\operatorname{tr} A A^\dag}
    \end{aligned}
\end{equation}
between an exact time evolution operator $U_{\mathrm{ex}}(t)$ and various approximations $U_{\mathrm{a}}(t)$. Here, $D_{\mathrm{dim}}$ is the dimensionality of the Hilbert space and was introduced such that for unitary time evolution, the mismatch would take values on the interval $[0,1]$.\\
Below in Fig. \ref{fig:compare} we compare the variational principle here (Eq. \ref{eq:action}) to a simple Taylor series
\begin{equation}
    e^{iHt}\approx 1-iHt-(tH)^2/2,
\end{equation}
a Kernel polynomial expansion ($\left\|.\right\|$ is the operator norm)
\begin{equation}
\begin{aligned}
     e^{-i Ht}\approx &\left[J_0(t\left\|H\right\|)-2J_2(t\left\|H\right\|)\right]\\
     &-2iJ_1(t\left\|H\right\|)\frac{H}{\left\|H\right\|}-4J_2(t\left\|H\right\|)\frac{H^2}{\left\|H\right\|^2},
\end{aligned}
\end{equation}
and another candidate variational principle defined by the action
\begin{equation}
    S=\int d t \operatorname{tr}\left[\left(i \partial_t U-H U\right)\left(i \partial_t U-H U\right)^\dagger\right]
\end{equation}

\begin{figure}[h]
    \centering
    \includegraphics[width=0.99\linewidth]{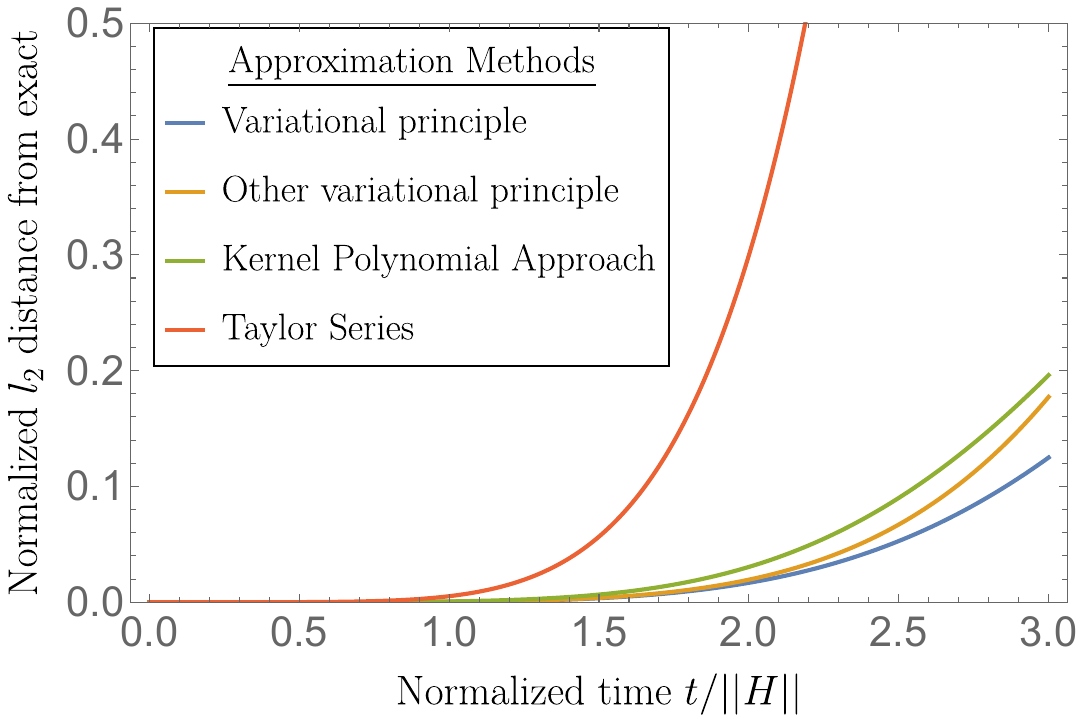}
    \includegraphics[width=0.99\linewidth]{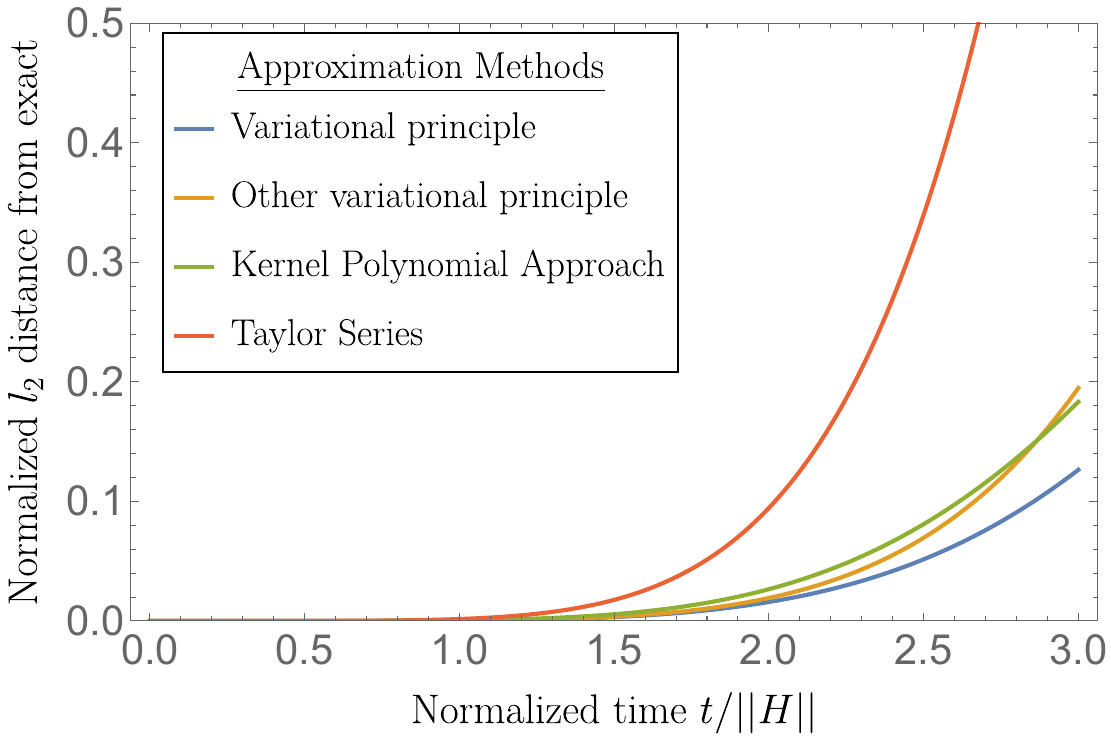}
    \caption{Presented is the $l_2$ distance between an exact time evolution operator and various approximate time evolution operators as a function of normalized time (normalized with the operator norm of the Hamiltonian). The curves averaged over 100 randomly chosen Hamiltonians. Here, the upper Plot corresponds to a  Hilbert space of dimension $d=5$, and the lower Plot to a Hilbert space of dimension $d=500$.}
        \label{fig:compare}
\end{figure}
The figure clearly shows that the results do not significantly depend on dimensionality - notably, the results did not differ significantly for most randomly chosen Hamiltonians in our test.

Perhaps it is useful to see an explicit form for differential equations 
\begin{equation}
    \begin{aligned}
        i  h_{i}\partial_t c_0+\sum_{j=1}^2h_{i+j} \left(i \partial_t c_{j}-c_{j-1}\right)-h_{i+3} c_{2}=0\\
        h_n=\mathrm{tr}(H^n)/D_{\operatorname{dim}};\quad \text{ with }i=0,1,2
    \end{aligned}.
    \label{eq:diffeqorder2}
\end{equation}
A simple, approximate, closed-form solution that is numerically very close to the exact solution for normalized times in the interval $[0,2]$ can be found by assuming the following form \footnote{Here, we used the typical initial conditions for coefficients and that $c_1$ can be found to have no quadratic term. The main additional approximation was setting $c_0=1$.}
\begin{equation}
\begin{aligned}
    c_0=1;\quad c_1=c_{13}t^3+c_{14}t^4-it;\quad c_2=c_{23}t^3+c_{24}t^4-\frac{t^2}{2}
\end{aligned}.
\end{equation}
A possible solution for the coefficients is then given as
\begin{equation}
    \begin{aligned}
        c_{13}&=\frac{i}{6}  \frac{h_3^2-h_2 h_4}{h_1 h_3- h_2^2};\quad c_{14}=\frac{1}{24}\frac{ h_3 h_4}{h_1 h_3-h_2^2}\\
        c_{23}&=\frac{i}{6}\frac{ h_1 h_4-h_2 h_3}{ h_1 h_3-h_2^2};\quad c_{24}=-\frac{1}{24}\frac{ h_2 h_4}{h_1 h_3-h_2^2}
    \end{aligned}.
\end{equation}
We note that the variational approach allowed us to decouple $t$ and $H$, unlike in a Taylor series where $tH$ always has to appear together. This additional flexibility in form is the root cause of the series' improvements. It allowed us to account for higher-order contributions of $H$ via traces of $H$ without significantly complicating the problem - actual operators only appear to order $ H^2$. 
\section{Application of the method to degenerate perturbation theory}
We next want to apply our approach to find non-perturbative improvements over degenerate second-order perturbation theory. The variant of degenerate perturbation theory that works best for our purposes uses unitary transformations. Consider a unitarily transformed Hamiltonian
\begin{equation}
   \scalemath{0.8}{ H^\prime =e^{-i[O,.]}(H_0+V)\approx H_0+V-i[O,H_0+V]-\frac{[O,[O,H_0+V]]}{2}},
\end{equation}
where $O$ is an anti-hermitian operator, $H_0$ is considered the unperturbed Hamiltonian and $V$ is treated as a perturbation. We also interpret the exponential as power series with $[O,.]^n A$ as nth nested commutator i.e. $[O,.]^2 A=[O,[O,A]]$ - that is it is the adjoined map. We will restrict our discussion to degenerate perturbation theory in a subspace with $\langle H_0\rangle=0$ and the case where $O$ can be chosen to fully remove the perturbation $V$ to lowest order, i.e., $V-i[O,H_0]=0$. If we set $V=\lambda V$ with $\lambda=1$ such that $\lambda$ can be used a formal expansion parameter, we realize $O\propto \lambda$. To second order in $\lambda$ we then obtain (setting $\lambda=1$)
\begin{equation}
    H^\prime \approx H_0-\frac{i}{2}[O,V],
\end{equation}
which only has to be projected onto the $\langle H_0\rangle=0$ subspace to obtain a simpler effective low energy Hamiltonian.\\
We note that $e^{-i[O,.]}$ contains non-perturbative information about the physical system that the typical degenerate perturbation theory does not capture. To extract this information without obtaining a complicated Hamiltonian, we define 
\begin{equation}
    \tilde O=(O\otimes \mathbb{1}-\mathbb{1}\otimes O^T)
\end{equation}
such that in a superoperator formalism \cite{superop,PhysRevLett.122.040604,Roberts_2024}, we may map 
$$e^{-i[O,.]}\to e^{-i\tilde Ot};\quad t=1$$
which can be expanded using the variational method that we discussed earlier. We find
\begin{equation}
    H^\prime \approx c_0(H_0+V)+c_1[O,H_0+V]+c_2[O,[O,H_0+V]]
    \label{eq:non-pertdegpert}
\end{equation}
with expansion coefficients that can be found by the previous method.\\
It is now useful to apply the method to example problems. The simplest example problem is AB bilayer graphene, which is described by
$$H_0=\gamma(\tau^+\otimes\sigma^++\tau^-\otimes\sigma^-);\quad V=\mathbb{1}_2\otimes \vect{p} \boldsymbol{\sigma},$$
where $\boldsymbol{\sigma}=(\sigma^1,\sigma^2)$ and $\sigma^i$ ($\tau^i$) are the typical Pauli matrices acting in sublattice (layer) space. We also made use of raising/lowering operators $\tau^\pm=\tau^1\pm i\tau^2$, $\sigma^\pm=\sigma^1\pm i\sigma^2$ and the vector of momentum operators $\vect p=(p_1,p_2)$. For the problem at hand, the operator $O$ that removes the perturbation $V$ to the lowest order is found very easily, making use of the superoperator formalism and a pseudoinverse as
\begin{equation}
    O=i([H_0,.])^{-1} V=-\frac{p_1\tau_2+p_2\tau_1}{\gamma}\otimes\sigma_3.
\end{equation}
The low-energy sub-space is spanned by $(0,0,1,0)$ and $(0,1,0,0)$, and our expression for ordinary second-order degenerate perturbation theory projected on this subspace then gives the result
\begin{equation}
    H^{(2)}=-\gamma^{-1}(p_+^2\sigma^++p_-^2\sigma^-);\quad p_\pm=p_1\pm p_2
\end{equation}
For the improved perturbation theory, we recognize that $V$, $H_0$, as well as $[O,[O,V]]$ vanish when projected onto the low energy subspace such that large parts of \eqref{eq:non-pertdegpert} collapse and one only has to compute
\begin{equation}
    H^{(2,\mathrm{var})}=\left(c_1-ic_2\right)[O,V]_{\mathrm{proj}}=2i\left(c_1-ic_2\right)H^{(2)}.
\end{equation}
The Hamiltonian is very similar to the typical degenerate second-order perturbation theory and is only corrected by a pre-factor. We now are in the position to find traces of $\tilde O$ using the shorthand notation $p=\sqrt{p_1^2+p_2^2}$ as
\begin{equation}
    \mathrm{tr}(\tilde O^{2n})=2^{n+2} \left[\left(\frac{p}{\gamma}\right)^n+\left(-\frac{p}{\gamma}\right)^n\right];\quad n\in \mathbb{N}^+
\end{equation}

This expression and Eq. \eqref{eq:diffeqorder2} allow us to write differential equations for the coefficients $c_i$. We find
\begin{equation}
    c_0=1;\quad c_1=-i\operatorname{sinc}\left(\frac{2 p}{\gamma}\right);\quad c_2=-\frac{1}{2}\operatorname{sinc}^2\left(\frac{p}{\gamma}\right)
\end{equation}
when we solve for coefficients $c_i$ at "time" $t=1$ and the final Hamiltonian then is given as
\begin{equation}
   H^{(2, \mathrm{var})}= \frac{\operatorname{sinc}^2\left(\frac{p}{\gamma}\right)-2\operatorname{sinc}\left(\frac{2p}{\gamma}\right)}{\gamma}\left(p_{+}^2 \sigma^{+}+p_{-}^2 \sigma^{-}\right)
\end{equation}
We note that the Hamiltonian has a rotationally invariant spectrum such that we can choose any direction and see how much the relative  mismatch 
\begin{equation}
    \Delta(\vect p) =\left|\frac{E_{\mathrm{ex}}(\vect p)-E_{\mathrm{app}}(\vect p)}{E_{\mathrm{ex}}(\vect p)}\right|
\end{equation}
between exact $E_{\mathrm{ex}}$ and approximate $E_{\mathrm{app}}$ energy levels changes as function of momentum. A plot is presented below in Fig. \ref{fig:ABbilayer_plot}.
\begin{figure}[h]
    \centering
    \includegraphics[width=1\linewidth]{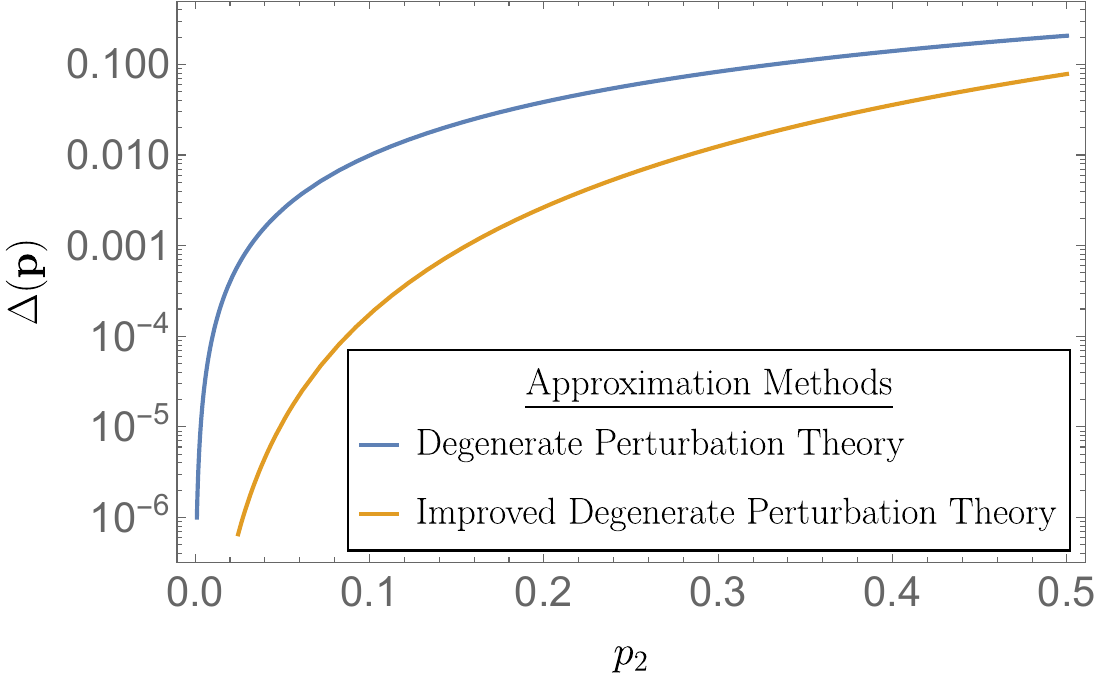}
    \caption{Plot of the relative mismatch between exact and approximate energies as a function of $p_2$ (the choice of direction does not matter because of rotational symmetry). The blue curve represents typical second-order degenerate perturbation theory, while the yellow curve corresponds to our improved methodology.}
    \label{fig:ABbilayer_plot}
\end{figure}
The figure shows that our new Hamiltonian is more reliable than typical degenerate perturbation theory - often by a few orders of magnitude.\\
This discussion raises the question of whether the approach can be used in a more useful context. One such context is mapping a Heisenberg model
\begin{equation}
    H=-t \sum_{\langle i j\rangle \sigma} c_{i \sigma}^{\dagger} c_{j \sigma}+U \sum_i n_{i \uparrow} n_{i \downarrow}
\end{equation}
to an effective spin model. Here, $c_i^\dag$ ($c_i$) are creation (annihilation) operators, and $n_i=c_i^\dag c_i$ the number operator for state $i$.

We will restrict ourselves to one dimension and apply the same approach as previously discussed in \cite{PhysRevB.37.9753}. That is, one first notes that an operator 
\begin{equation}
    O=-i\frac{t}{U}\sum_{\langle ij\rangle\sigma }(n_{i \bar{\sigma}} c_{i \sigma}^{\dagger} c_{j \sigma} h_{j \bar{\sigma}}-h_{i \bar{\sigma}} c_{i \sigma}^{\dagger} c_{j \sigma} n_{j \bar{\sigma}} )
\end{equation}
to first order removes terms coupling the half-filled subspace to the rest of the Hilbert space. Here, $\bar \sigma$ denotes a spin opposite of $\sigma$, and $h_i=1-n_i$ counts holes. With this choice of operator, we can follow an almost identical procedure as previously with AB bilayer graphene to find an effective Hamiltonian. In the case $N\geq 3$ sites, coefficients $c_i$ are found as 
\begin{equation}
\begin{aligned}
    &c_0=1;\quad c_1(t,U)=-i \operatorname{sinc}\left(\frac{\sqrt{\frac{3}{2}} t\sqrt{2 N+1} }{U}\right);\\
    &c_2(t,U)=-\frac{1}{2} \operatorname{sinc}\left(\frac{\sqrt{\frac{3}{2}} t\sqrt{2 N+1} }{2 U}\right)^2
\end{aligned}
\end{equation}
where $N$ is the number of sites in the system.\\
Finally, projecting on the half-filled subspace, we find a Heisenberg model with an improved energy scale
\begin{equation}
    H^{(2)}_{\mathrm{Var}}=-\frac{t^2}{U}\left[ic_1(t,U)+c_2(t,U)\right] \sum_{\langle i, j\rangle}\left(\mathbb{1}-\boldsymbol{\sigma}_i \cdot \boldsymbol{\sigma}_j\right)
\end{equation}
In the limit $t\to 0$, our result reduces to the well-known result
\begin{equation}
    H^{(2)}=\frac{-t^2}{2 U} \sum_{\langle i, j\rangle}\left(\mathbb{1}-\boldsymbol{\sigma}_i \cdot \boldsymbol{\sigma}_j\right)
\end{equation}
from typical degenerate perturbation theory. It is important to stress that, regrettably, higher-order corrections depend on $N$ explicitly in a way that is problematic in the thermodynamic limit $N\to \infty$. This challenge is due to the simplicity of the chosen ansatz that explicitly involves $1$,$O$, and $O^2$, which means the trace has contributions from different orders of $N$. Possible solutions could be to have a more careful ansatz that is local or a variational principle that is directly defined for the effective Hamiltonian. \\
Below in Fig. \ref{fig:Heisberg_approx}, we show plots that quantify the approximation error of the two approximate Heisenberg Hamiltonians $H^{(2)}_{\mathrm{Var}}$ (improved degenerate perturbation theory) and $H^{(2)}$ (ordinary degenerate perturbation theory). For simplicity, we considered the case of 5 sites\footnote{Other number of sites yield similar results, but filtering the spectrum according to states with half-filling and no double occupation becomes more cumbersome - less automatic}, periodic boundary conditions, and $U=1$.

\begin{figure}[h]
    \centering
    \includegraphics[width=1\linewidth]{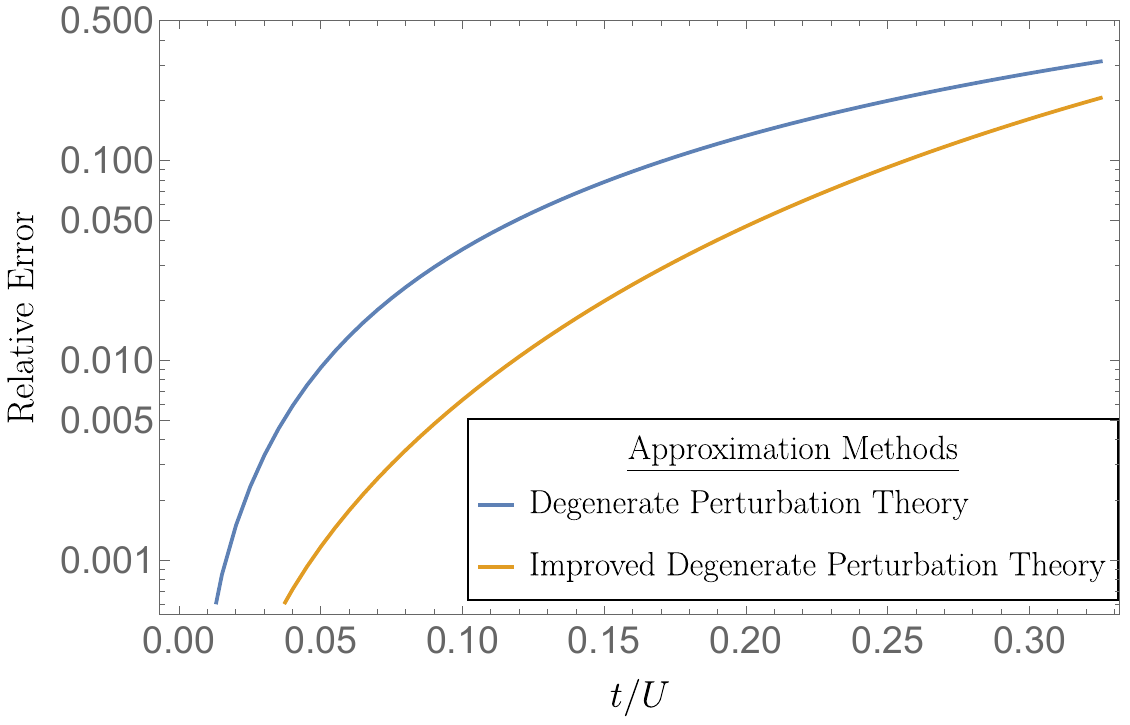}
     \includegraphics[width=1\linewidth]{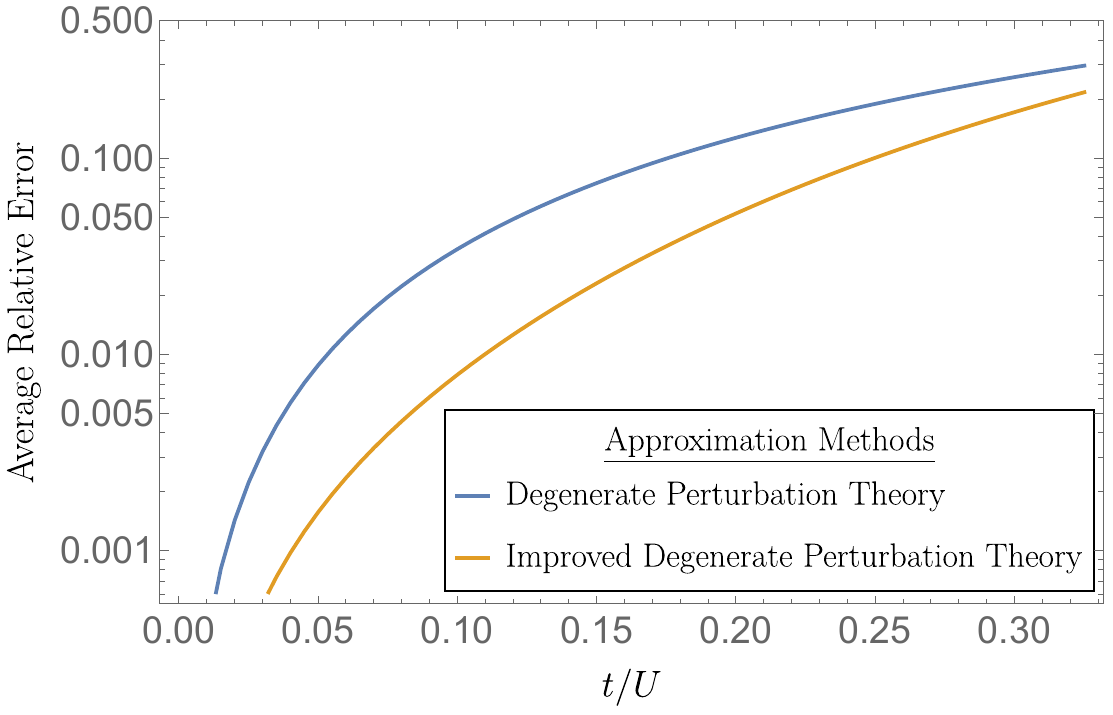}
    \caption{ The upper Plot shows the relative error $|(E_1^{\mathrm{ex}}-E_1^{\mathrm{app}})/E_1^{\mathrm{ex}}|$ between the lowest energy level of the approximate Hamiltonians $E_1^{\mathrm{app}}$ and the corresponding exact energy levels of the full Hubbard Hamiltonian $E_1^{\mathrm{ex}}$ as function of $t/U$ (for various of the other energy levels the plots are almost identical). The lower Plot averages the relative errors for the first half of non-zero energy levels (26 non-zero energy levels, and we consider the first 13 levels).}
    \label{fig:Heisberg_approx}
\end{figure}
The improved approach yields considerably better results for various energies while leaving wavefunctions unchanged. However, regrettably, this improvement is only present for the first half of non-zero energies - the other energies regrettably are not approximated as nicely as by typical degenerate perturbation theory. Moreover, we reiterate that the thermodynamic limit beyond the lowest order in $t$ is problematic, and the approximation becomes less reliable for huge numbers of sites $N$. However, up to our numerical capability $N=7$ (with the laptop at our disposal we did not have the numerical means to check for more than 7 sites since 8 sites for the full Hubbard model already has $(4^8)^2\approx$4 billion matrix elements) the improvements were quite nice. The issues with the thermodynamic limit can be traced to the appearance of different powers of $O$ in our ansatz and the corresponding traces. An approach that directly works with a variational principle for the effective Hamiltonian can be expected to provide a remedy and is the subject of future work in progress.

We also note that there are several opportunities for additional improvements. For instance, we have chosen the simplest first-order generator of unitary transformations - one that decouples the degenerate subspace of interest from the remaining Hilbert space to first order only - to obtain an effective Hamiltonian in the examples. Employing better generators has the potential to yield much further improvement.

\section{Conclusion}
\label{sec:concluscion}
We have demonstrated a variational method that allows for improved expansions of matrix exponentials. This approach is of particular interest when one is interest in finding reliable approximations to the time evolution operator of a quantum system. In the process, we found that expressions corresponding to a quantum geometric tensor for the time evolution operators naturally enter variational equations of motion. The key ingredient in the approach that permits improvements beyond a Taylor series $e^{-iHt}=\sum_{n=0}^{\infty} \frac{(-i H t)^n}{n!}$ - even when we expand in powers of $H$ - is that $t$ and $H$ in the resulting series are decoupled such that one can also have terms like $(c_1t+c_2t^3+\dots)H$ rather than just $tH$. This flexibility in form leads to more accurate approximations even at low orders in $H$. Beyond applications to time evolution, we have demonstrated that the approach can be usefully employed to improve degenerate perturbation theory. Here, we found that it did not complicate the operator content of resulting approximate Hamiltonians but improved the accuracy of energy levels. The approach gives a glimpse into the important role of quantum geometry in improving degenerate perturbation theory.\\
The idea in our work is general, and there is ample opportunity for deployment to different areas of theoretical physics and further generalization. For instance, the approach could be expanded to interacting quantum field theories (here, however, due to the unboundedness of bosonic operators, traces would have to be judiciously truncated, and one could work with Magnus expansions to adequately treat complications arising from an interaction picture) - in this case, it could yield a resummation of certain Feynman diagrams. Furthermore, the linear ansatz in powers of $H$ was the most naive choice; more judicious choices could yield further improvements. One could also imagine possible applications to Floquet theory, where one also uses unitarily transformed Hamiltonians. \\
In the most general context, we hope our work will inspire progress in understanding the connection between non-perturbative expansion methods and their underlying quantum geometry.
\section{Acknowledgements}
M.V. gratefully acknowledges the support provided by the Deanship of Research Oversight and Coordination (DROC) and the Interdisciplinary Research Center(IRC) for Intelligent Secure Systems (ISS) at King Fahd University of Petroleum \& Minerals (KFUPM) for funding his contribution to this work through internal research grant No. INSS2507. The author thanks Dr. Hocine Bahlouli for an insightful discussion of this manuscript.\\
 Data Availability Statement: No Data associated in the manuscript. Code used to generate figures can be shared upon reasonable request.

\bibliographystyle{unsrt}
\bibliography{ref}
\end{document}